\documentclass[aps,prl,twocolumn,showpacs,floatfix,preprintnumbers,amsmath,amssymb]{revtex4}
\usepackage{epsfig}
\usepackage{graphicx}
\usepackage{dcolumn}
\usepackage{bm}

\begin{document}
\title{Spectral fluctuations and 1/f noise in the order-chaos transition regime}
\author{M. S. Santhanam and Jayendra N. Bandyopadhyay}
\affiliation{Physical Research Laboratory,
Navrangpura, Ahmedabad 380 009, India.}

\begin{abstract}
Level fluctuations in quantum system have been used to characterize
quantum chaos using random matrix models. Recently
time series methods were used to relate level fluctuations
to the classical dynamics in the regular and 
chaotic limit. In this we show that the spectrum of the system undergoing
order to chaos transition displays a characteristic
$f^{-\gamma}$ noise and $\gamma$ is correlated with the classical
chaos in the system. We demonstrate this 
using a smooth potential and a time-dependent system modeled by Gaussian and 
circular ensembles respectively of random matrix theory. We show
the effect of short periodic orbits on these fluctuation measures.
\end{abstract}

\maketitle

Quantum chaos, the study of quantum analogues of classically
chaotic systems, is characterized by the fluctuation
properties of the spectrum of its Hamiltonian operator.
For quantum systems
with regular classical dynamics the spectral fluctuations
are Poisson distributed \cite{bt}, i.e,
the eigenvalues tend to cluster together. On the other hand,
one of the remarkable results established by Bohigas {\it et al}
is that the level fluctuation properties of quantum systems,
whose classical limit is chaotic, are identical to
those of an appropriate ensemble from random matrix theory (RMT) \cite{boh1}.
This is the level repulsion regime where the eigenvalues tend to
repel one another.
In this sense, characterizing quantum chaos in terms of
presence or absence of level repulsion requires invoking the
spectral properties of random matrix ensembles. Recently, in analogy
with time series, a method
has been proposed to characterize spectral fluctuations using
inherent properties of the spectrum \cite{rel1}.
If the eigenvalues of Hamiltonian operators could be thought
of as a time series and its index
the time in some units, then the methods of traditional time series
analysis can be applied to it. It was shown that
the ensemble averaged power spectrum $\langle S(f) \rangle$ of the
fluctuations in the cumulative level density,
goes as $1/f$ or $1/f^2$ depending on
whether the system is classically chaotic or regular \cite{rel2}.
This work also showed examples of atomic level sequences displaying
$1/f$ noise. Thus, atomic levels join the host of other systems and phenomena
that display $1/f$ noise lending strength to the well known clich\'{e} that
$1/f$ noise is ubiquitous in nature \cite{obf}.

In this paper, we study the transition from
regularity to chaos in the mixed systems. In such systems
the regular and chaotic motion coexist and this is a generic
feature. For instance,
the entire class of atoms in strong fields and the range of problems
involving atoms in time-varying fields
belong to this class. We study a smooth Hamiltonian
system, the quartic oscillator and a time dependent system,
the kicked top. In both these systems, a single parameter
that controls the classical chaos
can be varied to get a smooth
transition from regular to predominantly chaotic dynamics.
It is well known that the level fluctuations in these systems can
be modelled by RMT \cite{boh1}.
From an RMT point of view, these models possess symmetries
(time-reversal invariance without spin-1/2 interactions) such that
the kicked top is part of the
circular orthogonal ensemble whereas the coupled oscillator
falls in the Gaussian orthogonal ensemble (GOE) of RMT \cite{rmt}.
Based on the numerical evidence from these models we show that
$\langle S(f)\rangle \; \varpropto f^{-\gamma}$,
where $\gamma$ depends on the degree of their classical chaos.
This correlation between $\gamma$ and the classical chaos parameter
is established using semi-empirical level spacing distributions studied
in the context of RMT to model the transition region.

In semiclassical systems of the type we consider here, periodic
orbits via the Gutzwiller formalism play an important role in
determining the quantum spectrum \cite{rmt}. As pointed out by
Berry \cite{ber1}, the properties of the spectrum on a scale of mean level
spacing are determined by long time period orbits and in a sense
both of them display universality; the RMT type universality in the
spectrum and classical universality embodied in the Hannay-Ozorio sum
rule \cite{hor}. However, long range spectral properties are determined by
the short time periodic orbits which are system specific and are 
not universal. This manifests itself in the power spectrum as deviations
from the $f^{-\gamma}$ behavior.
We show the effect of short time periodic orbits in the
coupled oscillator, where scarring or the density enhancements
in the vicinity of certain periodic orbits \cite{hell} is a prominent feature
due to these orbits.

The Hamiltonian for the coupled quartic oscillator is,
\begin{equation}
H_{1} = p^{2}_{1} + p^{2}_{2} +q_1^{4}+q_2^{4}+\alpha ~ q_1^{2} q_2^{2}.
\label{qosc}
\end{equation}
This system is classically integrable for $\alpha=0, 2, 6$ and
the phase space is predominantly chaotic for $\alpha > 6$.
This has been extensively studied as a model for chaotic dynamics in
a smooth potential \cite{qo,eck}. The Hamiltonian is quantized by
solving the corresponding Schr\"{o}dinger equation
in the basis of the eigenfunctions
corresponding to $\alpha=0$. Then, the matrix elements of
Hamiltonian operator $\widehat H_1$ is computed in this
basis and the matrix of order 13000 is
diagonalized to obtain about 2000 converged eigenvalues.
This system possesses $C_{4v}$ point group symmetry.
We symmetry decompose the spectrum and
study the levels from only the  $A_1$  irreducible
representation.

The quantum top \cite{top1} is characterized by an angular momentum
vector ${\bf J}$, whose components $(J_x,J_y,J_z)$ obey the
usual commutation relations and ${\bf J^2}=j(j+1), j=\frac{1}{2}
,1,\frac{3}{2}$,......, is conserved.
The dynamics of the top is governed by the Hamiltonian \cite{top1},
$H_{2}(t) = \frac{\pi}{2} J_y + \frac{k}{2 j} J_z^2 \sum_{n}
\delta(t-n)$.
The first term describes the precession around the $y$ axis with
angular frequency $\pi/2$ and the second term kicks in periodically
with $\delta$-function kicks of strength $k$. Each kick can be thought of as an
impulsive rotation about $z$-axis by an angle $ k J_z/2 j$.
The time evolution operator in between consecutive kicks
is,
\begin{equation}
U = \exp\left(-i \frac{k}{2j} J_z^2 \right) \exp\left(-i \frac{\pi}{2} 
J_y\right).
\label{uop}
\end{equation}
If $e^{i \phi}$ are the eigenvalues of $U$, then the power spectrum
is computed from the quasienergy $\phi$. In the limit $j\rightarrow \infty$,
we can derive a classical map whose dynamics depends on
the parameter $k$ \cite{top1}.  At $k=0$ it is integrable
and becomes increasingly chaotic for $k > 0$.

We denote the eigenvalues of the appropriate operator
described above by $E_i$, $i=1,2,\dots,n+1$.
The integrated level density, that counts the number of levels below
a given $E$, can be decomposed into an average and an oscillating
part, $N(E) = \overline{N}(E) + N_{osc}(E)$.
In order to compare the fluctuations from various systems it is customary
to unfold the spectrum by a transformation
$\lambda_i = \overline{N}(E_i)$, such that the mean level density of the 
transformed levels is unity. All further analysis is 
carried out using the sequence $\{\lambda \}$.
For instance, the spacing is
$s_i = \lambda_{i+1} -  \lambda_i, i=1,2,\dots,n$.
In this paper, we will work with the statistic given by,
\begin{equation}
\delta_m = \sum_{i=1}^{m} (s_i - \langle s \rangle ) \equiv - N_{osc}(E_{m+1})
\;\;\;\;\;\; m=1,2,...n
\label{dn}
\end{equation}
Once the unfolding is performed, on an average, a unit interval
of the spectrum will have one level. Hence,
$\delta_m$  represents the cumulative deviation until $m$th level, of the $i$th unfolded
level from $i$. This quantity has a formal analogy with
a time series. If the index $i$ represents the (scaled) time,
then $(s_i - \langle s\rangle)$ represents the actual value assumed by the series 
at the $i$th time instant. 
Following Ref. \cite{rel1}, we take the power spectrum of $\delta_m$ as,
\begin{equation}
S(f) = | \widehat\delta_f|^2  = \left| \frac{1}{\sqrt{n+1}} \sum_m \delta_m \exp\left( \frac{-2 \pi i f m}{n+1} \right) \right|^2
\label{pspect}
\end{equation}
where $f=1,2,\dots,n ;\,\, \widehat\delta_f$ is the Fourier transform of 
$\delta_m$.  Now, we will present results to infer that
$ \langle S(f) \rangle  \varpropto f^{-\gamma}$
and $\gamma$ depends on the degree of chaos in the system.

Before we plunge into the results, we show the statistic $\delta_m$ for
our models.
For the quartic oscillator, using the results
in Refs. \cite{mart}
we symmetry decompose the level density
to obtain the asymptotic
integrated level density for the $A_1$ representation,
\begin{equation}
\begin{split}
\overline{N}_{A_1}(E) & \approx  \frac{1}{8} \left[ \frac{E^{3/2}}{6} 
F\left(\frac{1}{2},\frac{1}{2};1; \frac{2-\alpha}{4}\right)   \right. \\
+& \left.
\frac{\Gamma(5/4) E^{3/4}}{\Gamma(7/4) \sqrt{\pi} 2^{7/4}} +
\frac{E^{3/4}}{\Gamma(7/4) \sqrt{2\pi} (2+\alpha)} +  \frac{3}{4} 
\right]
\end{split}
\label{idos_qosc_a1}
\end{equation}
where $F(.)$ is the Gauss' Hypergeometric function.
We use this expression to unfold the quartic oscillator levels.
\begin{figure}
\includegraphics[height=6cm,width=7cm]{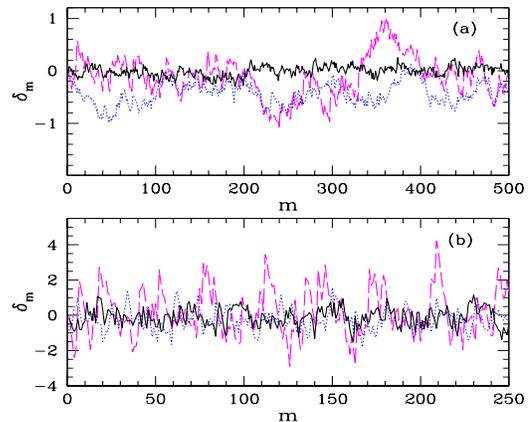}
\caption{(Color online) $\delta_m-m$ curve for (a) kicked top and (b) quartic 
oscillator. The solid curve corresponds to chaotic case $(k=7, \alpha=20)$ 
and the dotted curve
to the intermediate region $(k=3, \alpha=12)$,
and the dashed curve to regular limit $(k=1,\alpha=0)$.}
\label{dn-n}
\end{figure}
Fig. \ref{dn-n} displays $\delta_m - m$ for the quartic 
oscillator
and the kicked top for a choice of 3 parameters in the regular, chaotic
and the transition region. In Fig. \ref{dn-n}(a) for the kicked top, the dashed
curve corresponding to nearly the Poisson spectrum ($k=1$) differs markedly from
the solid curve for almost the GOE $(k=7)$ limit.
We also plot an intermediate case (dotted curve) to show the transition taking 
place from Poisson to GOE type spectrum.
This intermediate case at $k=3$ is also qualitatively different. Each series
displays slightly different memory effects corresponding to various shades
of anti-persistent time series. In fact, such time series are known
to display $f^{-\gamma}$ noise and we expect similar result
based on this analogy.
We observe similar features for the quartic oscillator in Fig. \ref{dn-n}(b)
for $\alpha=6, 11.5, 19.5$,
as reported in an earlier work of Bohigas \cite{boh2}.

\begin{figure}
\includegraphics[height=7cm]{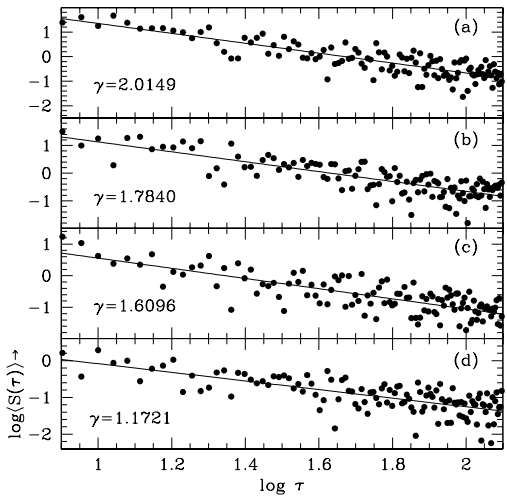}
\caption{Power spectrum of $\delta_n$ for kicked top at
(a) $k=1$, (b)$k=2.0$, (c) $k=3.0$ and (d) $k=7.0$. The solid
line is the least squares fit. The slope $\gamma$ is indicated in each graph.}
\label{ps-top}
\end{figure}

\begin{figure}
\includegraphics[height=7cm]{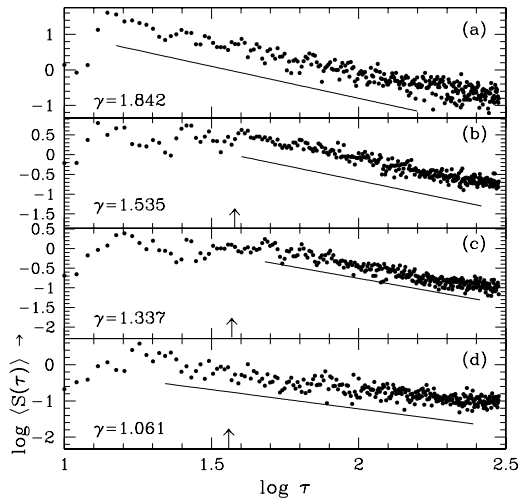}
\caption{Power spectrum of $\delta_m$ for quartic oscillator at (a) $\alpha$=0,
2,6 (b) $\alpha$=7.5 (c) $\alpha$=11.5 and (d) $\alpha$=19.5 . The solid lines 
are the least squares fit with intercept shifted for clarity. The slope 
$\gamma$ is indicated in each graph. The arrows in (b-d) indicate
$\log f_{\mbox{\small min}}$, the time period of the short periodic orbit. }
\label{ps-qosc}
\end{figure}

In Figs. \ref{ps-top} and \ref{ps-qosc}, we display the ensemble
averaged power spectrum of $\delta_m$.
Ensemble averaging is done as follows :
for each $\alpha$ of the quartic oscillator we obtain
2000 levels. After leaving out the first 200 levels, we create
3 sequences of 600 levels each. We further obtain similar
sequences from more values of $\alpha$ separated by $\delta\alpha=0.1$. For 
instance, $\alpha=11.5$ in Fig. \ref{ps-qosc}(b)
corresponds to ensemble average in the range $\alpha=11.2-11.8$ in steps of 
0.1. Hence, the results (except for the integrable case in Fig. \ref{ps-qosc}(a))
represent an average over 21 level sequences of length 600 each.
For the integrable case, the ensemble consists 9 level sequences (3 from each
of $\alpha=0,2$ and 6) of length 600
each. Similar averaging is done for the kicked top with 11 level
sequences of length 800 each.
For the regular and chaotic limits shown in
Figs. \ref{ps-top}(a,d) and \ref{ps-qosc}(a,d),
there is a good agreement with the predicted slopes of 
$\gamma =2,1$ respectively \cite{rel2}.
There are deviations for $f < f_{\mbox{\small min}}$
due to the effect of short periodic orbits with scaled
time period $f_{\mbox{\small min}}$. The deviation
for large $f$ arises, partly, from approaching the Nyquist frequency at
$f = (n+1)/2$. It is clear from Figs. \ref{ps-top},\ref{ps-qosc}(b,c) 
that as the oscillator and the top explore the intermediate region between
the regular and the chaotic limits, the slope
$\gamma$ smoothly changes from 2 to 1.
Intuitively, we can expect this because other
statistic' in the transition regime, e.g,
spacing distribution, generally vary smoothly too in this regime.

\begin{figure}
\includegraphics[height=6cm,width=7cm]{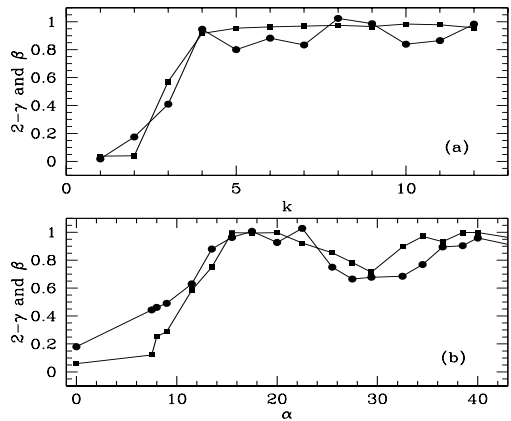}
\caption{The parameter $\beta$ (squares) in $P_T(s)$ as a function of chaos parameter
for (a) the kicked top and (b) quartic oscillator. Also plotted is the
shifted slope $2-\gamma$ (circles) of the power spectrum $\langle S(f) \rangle$.
The slope is  shifted such that $y-$axis lies in the range 0 to 1.}
\label{brody}
\end{figure}

In order to obtain a global picture, we compute the exponent $\gamma$ for
a range of parameter values of $\alpha$ and $k$ in the order-chaos
transition region.
In RMT, this transition is described by a semi-empirical
spacing distribution $P(s;\beta)$ characterised by the parameter $\beta$.
As $\beta$ is varied from 0 to 1, $P(s;\beta)$ changes smoothly from 
Poisson to GOE, reflecting the change
in classical dynamics from regularity to chaos. Here,
we use the distribution $P_T(s;\beta)$ due to Izrailev \cite{izr}.
For $\beta=0$, $P_T(s;\beta)$ gives a Poissonian form (integrable limit)
and for $\beta=1$ it closely approximates the GOE distribution (chaotic limit).
The intermediate values, $0 < \beta < 1$, correspond
to order-chaos transition. Fig. \ref{brody}(a,b) shows that the series of
$\beta$ and $\gamma$ display similar trends and are strongly correlated for both
the top and the oscillator. It is known that the fraction of regular regions
in phase space is correlated with a parameter like $\beta$ that characterizes
the change in spacing distribution \cite{zimm}.
Hence we infer that the
exponent $\gamma$ in the power spectrum reflects the
qualitative trends in the classical dynamics of the system.
In general, we have shown that
$\langle S(f) \rangle \varpropto f^{-\gamma}$, where the value
of $1 \le \gamma \le 2$
relates to the nature of classical dynamics for the oscillator and the top.

As opposed to purely random matrices,
the semiclassical systems deviate from $f^{-\gamma}$ scaling \cite{rel2}
for $f < f_{\mbox{\small min}}$, where
$f_{\mbox{\small min}}$ is the period $t_{\mbox{\small min}}$ of the 
shortest periodic orbit (PO) scaled by the Heisenberg time $(t_H=2\pi\hbar)$ \cite{ber1},
i.e, $f_{\mbox{\small min}} = (n+1) t_{\mbox{\small min}}/t_H$.
This corresponds to short POs being
system specific features and in the corresponding large energy scales
universality breaks down leading to deviations from RMT based results \cite{ber1}.
Since the spectral form factor is linear only for times $t/t_H < 1$, which is necessary
to realise $1/f$ noise \cite{rel2}, the actual range of scaling is restricted
to $f_{\mbox{\small min}} \ll f \ll f_H$.
Since all times are scaled by $t_H$, at $t=t_H$,
we have $f_H=(n+1)$.
The short PO in the quartic oscillator is the
`channel orbit', as it is referred to in the literature,
with the initial condition $(q_1=0,q_2,p_1=0,p_2)$. 
This can be identified by taking discrete Fourier transform of
the scaled energies of the oscillator \cite{boh2}.
The period of this orbit is
$t(E) = E^{-1/4} \frac{\sqrt{\pi} \Gamma(1/4)}{2 \; \Gamma(3/4)}$.
Among all the sequence of levels that form
the ensemble let the largest level be $E=E'$. Then, the
period of the short PO is $t_{min} = t(E')$.
For instance, at
$\alpha=11.5$, we have $E'=2579.05$ ($\hbar=1$) and $n=600$ and this provides
the theoretical bounds for scaling to be $\log(f_{min})=1.54$ and $\log(f_H)=2.77$,
as indicated in Fig. \ref{ps-qosc}(c).
It is evident from Fig. \ref{ps-qosc}(b-d) that
$\langle S(f) \rangle \varpropto f^{-\gamma}$ in almost the entire 
theoretically expected range $f_{\mbox{\small min}} \ll f \ll f_H$.
This scaling range can be increased by probing deep in the semiclassical ($\hbar \to 0$)
regime, since $t_H \backsim \hbar^{-1} \backsim E^{3/4}$ for the quartic oscillator. But this 
leads to large eigenvalue problems that may not be computationally
feasible at present.

\begin{figure}
\includegraphics[height=5cm,width=7cm]{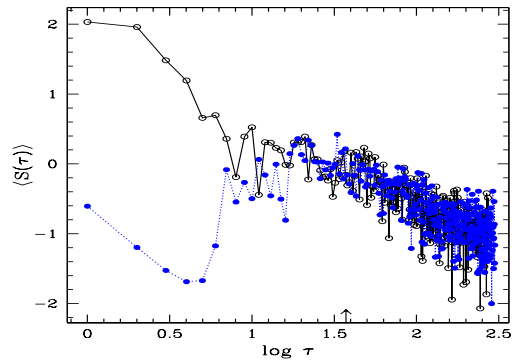}
\caption{(Color online) The effect of short periodic orbits in the spectrum.
$\langle S(f) \rangle$ for quartic oscillator at $\alpha=30$ including the
localized states (dotted). The arrow indicates
$f_{\mbox{\small min}}$. The solid line is the power spectrum after
localized states are removed.}
\label{porb}
\end{figure}

In Fig. \ref{porb}, we show the effect of short periodic orbits
on the power spectrum of $\delta_m$. 
Note that for the quartic oscillator at $\alpha=20$, $\gamma=1.073$
whereas for $\alpha=30$  we have $\gamma=1.322$ and $\beta$ also
shows a similar trend (see Fig \ref{brody}). In general, we would
expect that as $\alpha$ increases monotonically, chaos also increases
and hence agreement with RMT should get better. But numbers quoted
above show that, roughly speaking, chaos at $\alpha=20$ is
more than at $\alpha=30$.  This `anomalous' feature
is the effect of oscillating stability of the
short PO. At $\alpha=20, 30$ the short PO undergoes an anti-pitchfork and
a pitchfork bifurcation respectively accompanied by local
changes in the phase space structure \cite{mss2}. This, in turn,
affects the spectral 
levels.  It is known that the short POs influence a series of 
eigenstates, called the localized or sometimes the scarred states, in the
quartic oscillator spectrum and they deviate strongly
from RMT for eigenvector statistics \cite{eck,san,mss}.
If we remove the spacings that involve localized states, then we
might expect the resulting distribution to show a better agreement with RMT.
This is like removing the effect of short POs in the spectrum. 
The dotted line in Fig. \ref{porb} is the usual power spectrum and the
solid line is the one whose spacings involving localised
states are removed.  In this example the
ensemble has just 3 sequences of 600 levels each which is
reflected in large amplitude of fluctuations.
The power spectrum
changes character for $f < f_{\min}$ and the range of
validity of power law gets better. This is a manifestation
of the effect of short POs in the spectrum.

In summary, we have shown that the spectral fluctuations in the
quartic oscillator and the kicked top display $f^{-\gamma}$ 
noise, where the value of $\gamma$ within
the limits $1 \le \gamma \le 2$ reflects the
underlying nature of classical dynamics, namely, regular or
chaotic or a mixture thereof. We show the effect of short POs
on the spectral fluctuations.
We expect $f^{-\gamma}$ type noise in the
level fluctuations to be an inherent characteristic of quantum systems.

\acknowledgments
After this work was completed, qualitatively similar results obtained
on a billiard system were posted on LANL archive \cite{rel3}.
We thank Prof. V. B. Sheorey and Dr. Dilip Angom for useful discussions.

\end{document}